\documentclass[twocolumn]{emulateapj}
\usepackage{apjfonts}

\slugcomment{Accepted for publication in ApJ}

\shorttitle{Central BSSs in four outer-halo GCs}
\shortauthors{Beccari et al.}

\begin{document}

%% LaTeX will automatically break titles if they run longer than
%% one line. However, you may use \\ to force a line break if
%% you desire.

\title{The central Blue Straggler population in four outer-halo globular clusters}

%% Use \author, \affil, and the \and command to format
%% author and affiliation information.
%% Note that \email has replaced the old \authoremail command
%% from AASTeX v4.0. You can use \email to mark an email address
%% anywhere in the paper, not just in the front matter.
%% As in the title, use \\ to force line breaks.

\author{Giacomo Beccari\altaffilmark{1},
Nora L\"{u}tzgendorf\altaffilmark{1},
Christoph Olczak\altaffilmark{2,3,4},
Francesco R. Ferraro\altaffilmark{5},
Barbara Lanzoni\altaffilmark{5},
Giovanni Carraro\altaffilmark{6,7},
Peter B. Stetson\altaffilmark{8},
Antonio Sollima\altaffilmark{9}\and
Henri M.J. Boffin\altaffilmark{6}}

%% Notice that each of these authors has alternate affiliations, which
%% are identified by the \altaffilmark after each name.  Specify alternate
%% affiliation information with \altaffiltext, with one command per each
%% affiliation.

\altaffiltext{1}{European Southern Observatory, Karl-Schwarzschild-Strasse 2, 85748 
Garching bei M\"unchen, Germany}
\altaffiltext{2}{Astronomisches Rechen-Institut (ARI), Zentrum f\"ur Astronomie Universit\"at Heidelberg, 
M\"onchhofstrasse 1214, 69120 Heidelberg, Germany }
\altaffiltext{3}{Max-Planck-Institut fr Astronomie (MPIA), K\"onigstuhl 17, 69117 Heidelberg, Germany}
\altaffiltext{4}{National Astronomical Observatories of China, Chinese Academy of Sciences (NAOC/CAS), 20A 
Datun Lu, Chaoyang District, Beijing 100012, PR China}  
\altaffiltext{5}{Dipartimento di Astronomia, Universit\`a degli Studi di Bologna, via Ranzani 1, I-40127 Bologna, Italy}
\altaffiltext{6}{European Southern Observatory, Alonso de Cordova 3107, Santiago de Chile, Chile}
\altaffiltext{7}{Dipartmento di Astronomia, Universit\'a di Padova, Vicolo Osservatorio 3, I- 35122, Padova, Italy}
\altaffiltext{8}{National Research Council of Canada, Herzberg Institute of Astrophysics, 5071 West Saanich Road, Victoria, BC V9E 2E7, Canada}
\altaffiltext{9}{INAF - Osservatorio Astronomico di Padova, Vicolo dell'Osservatorio 5, 35122 Padova, Italy}

\begin{abstract}
Using HST/WFPC2 data, we have performed a comparative study of the
Blue Straggler Star (BSS) populations in the central regions of the
globular clusters AM 1, Eridanus, Palomar 3, and Palomar 4.  Located
at distances $R_{GC} > 50$ kpc from the Galactic Centre, these are
(together with Palomar 14 and NGC 2419) the most distant clusters in
the Halo.  We determine their colour-magnitude diagrams and centres of
gravity.  The four clusters turn out to have similar ages (10.5-11
Gyr), significantly smaller than those of the inner-Halo globulars,
and similar metallicities.  By exploiting wide field ground based
data, we build the most extended radial density profiles from resolved
star counts ever published for these systems.  These are well
reproduced by isotropic King models of relatively low concentration.
BSSs appear to be significantly more centrally segregated than red
giants in all globular clusters, in agreement with the estimated {\bf
  core} and half-mass relaxation times which are smaller than the
cluster ages.  Assuming that this is a signature of mass segregation,
we conclude that AM 1 and Eridanus are slightly dynamically more
evolved than Pal 3 and Pal 4.
  
 \end{abstract}

%% Keywords should appear after the \end{abstract} command. The uncommented
%% example has been keyed in ApJ style. See the instructions to authors
%% for the journal to which you are submitting your paper to determine
%% what keyword punctuation is appropriate.


%% From the front matter, we move on to the body of the paper.
%% In the first two sections, notice the use of the natbib \citep
%% and \citet commands to identify citations.  The citations are
%% tied to the reference list via symbolic KEYs. The KEY corresponds
%% to the KEY in the \bibitem in the reference list below. We have
%% chosen the first three characters of the first author's name plus
%% the last two numeral of the year of publication as our KEY for
%% each reference.

%% Authors who wish to have the most important objects in their paper
%% linked in the electronic edition to a data center may do so by tagging
%% their objects with \objectname{} or \object{}.  Each macro takes the
%% object name as its required argument. The optional, square-bracket 
%% argument should be used in cases where the data center identification
%% differs from what is to be printed in the paper.  The text appearing 
%% in curly braces is what will appear in print in the published paper. 
%% If the object name is recognized by the data centers, it will be linked
%% in the electronic edition to the object data available at the data centers  
%%
%% Note that for sources with brackets in their names, e.g. [WEG2004] 14h-090,
%% the brackets must be escaped with backslashes when used in the first
%% square-bracket argument, for instance, \object[\[WEG2004\] 14h-090]{90}).
%%  Otherwise, LaTeX will issue an error. 

\section{Introduction}

In the colour-magnitude diagram (CMD) of a globular cluster (GC), Blue
Straggler Stars (BSSs) define a sub-population located along the
main-sequence (MS) in a position brighter and bluer than the current
MS-turnoff (TO). For this reason these objects are thought to be
hydrogen burning stars more massive than a typical TO star. Two
  physical mechanisms are thought to be responsible for their
  formation: mass-transfer in a binary system \citep{mc64}, and direct
  stellar collisions \citep{hi76}.  Since collisions are more frequent
  in regions of higher density, the relative importance of the
  different formation channels could depend on the environment
  \citep[e.g., ][]{fu92,dav04}: BSSs in loose GCs might preferentially
  arise from mass-transfer activity in primordial binaries
  \citep[hereafter MT-BSSs;][]{fe06, le11a}, while BSSs located in high
  density environments might mainly form from stellar collisions
  \citep[hereafter COL-BSSs;][]{fe04}.

Discovered for the first time by~\citet[][]{sa53} in the external
regions of the GC M3, it was with the advent of the Hubble Space
Telescope (HST) and the 8-meter class telescopes that it became
possible to search for BSSs in dense cluster cores
\citep[e.g.][]{par91,fepar93,cl04}. BSSs have been found in any
properly imaged GC \citep{pi04} and they are studied also in dwarf
spheroidal galaxies \citep{ma07,mon12}.

Since the BSS formation mechanisms seems to be tightly connected with the
cluster internal dynamics, these stars are commonly recognized as ideal test particles to investigate the
impact of dynamics on stellar evolution in different environmental
conditions \citep[e.g.][]{ba95, si99, mo08, felan09}.  An interesting
example comes from the discovery of two distinct sequences of BSSs in
the core of M30 \citep{fe09}.  The authors argued that each of the two
sequences is populated by BSSs originated by one of the two formation
channels, both triggered by the collapse of the cluster core a few Gyr
ago. On the other side, ~\citet[][]{kn09} suggest that most BSSs, 
even those found in cluster cores, come from binary systems~\citep[see also][]{le11b}. 
Nevertheless the parent binaries may themselves have been affected by dynamical encounters.

Systematic studies of the BSS populations in GCs have
shown that in most cases their radial distribution is bimodal, with a
high peak in the cluster centre, a trough at intermediate radii and a
rising branch in the external regions \citep[see e.g.][ and references
  therein]{da09}.  Dynamical simulations \citep[e.g.][]{si94, ma04,
  ma06, la07a,la07b} suggest that the observed shape of the BSS radial
distribution depends on the relative contribution of COL-BSSs and the
efficiency of the dynamical friction, that progressively segregates
objects more massive than the average (as BSSs and their progenitors)
towards the cluster centre. Very interesting exceptions are the cases
of $\omega$ Centauri \citep{fe06}, NGC 2419 \citep{da08} and Pal 14
\citep{bec11}, where the BSS radial distribution is completely flat.
This fact indicates that in these clusters the dynamical friction was
not effective yet in segregating BSSs toward the centre of the
potential well.  Interestingly enough NGC 2419 and Pal 14 are among
the most remote GCs in the Galaxy, lying at a distance from the
Galactic centre $R_{GC} > 50$ kpc.

We decided to extend the investigation to other four Galactic GCs
(namely AM 1, Eridanus, Pal 3, Pal 4) located in the extreme
outer-halo.  In Section \ref{sec_data} we describe the data-set and
reduction procedure. The CMD and the age estimate for each cluster are
discussed in Section \ref{sec_cmd}. The determination of the
astrometric solutions and the cluster centres of gravity is presented
in Section \ref{sec_astro}.  The radial density profiles from resolved
star counts and the estimates of the cluster structural parameters and
characteristic time-scales are discussed in Section
\ref{sec_dens}. Section \ref{sec_bss} is devoted to the radial
distribution of the BSS, while a summary is presented in Section
\ref{sec_summ}.

%%%%%%%%%%%%%%%%%%%%%%%%%%%%%%%%%%%%%%% SECTION
%%%%%%%%%%%%%%%%%%%%%%%%%%%%%%%%%%%%%%% FIGURE
\begin{figure}
\centering
\includegraphics[scale=0.44]{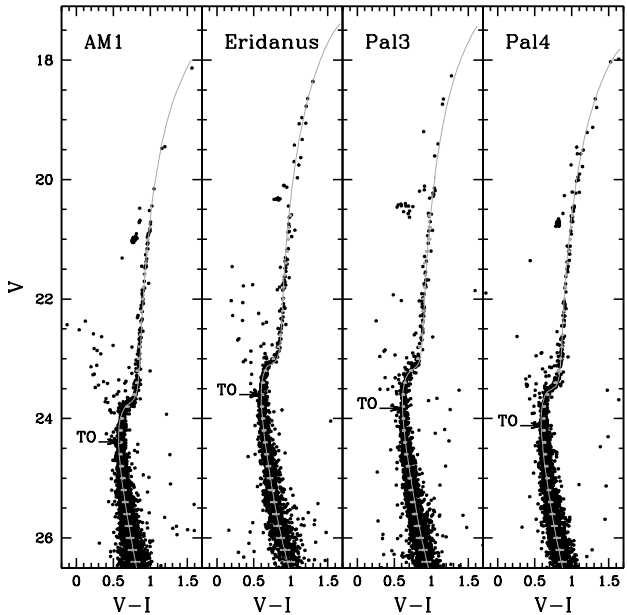}
\caption{(V,V-I) CMDs of the GCs AM 1, Eridanus, Pal 3, Pal 4. All the
  sampled stars are shown with black dots, while the location of the
  mean ridge line is shown with grey points. We fitted the CMD with
  isochrones from \citet[][grey solid lines]{do08b}. The location of
  the MS-TO is also indicated.}
\label{fig_cmd}
\end{figure}
%%%%%%%%%%%%%%%%%%%%%%%%%%%%%%%%%%%%%%% FIGURE

\section{The data set}
\label{sec_data}
\citet[][ hereafter S99]{ste99} published deep optical CMDs obtained
with the Wide Field Planetary Camera 2 (WFPC2) on board the HST for
Eridanus, Pal 3 and Pal 4. The core region of Eridanus was roughly
centred on the WF3 chip of the WFPC2 mosaic, while Pal 3 and Pal 4
were centred on the PC chip. We refer to S99 \citep[see also][]{ha97}
for a detailed description of data quality and reduction procedure. In
brief, a standard DAOPHOTII \citep[][]{ste87} Point Spread Function
(PSF) fitting procedure, including ALLFRAME \citep[][]{ste94}, was
adopted to obtain instrumental magnitudes and colours for all
measurable stars in the WFPC2 fields.  Here we adopted the S99
photometric catalogues of Pal 3, Pal 4 and Eridanus listing the
$F555W$ and $F814W$ magnitudes of 2968, 4122 and 2340 stars,
respectively.

AM 1 was observed with the WFPC2 in the $F555W$ and $F814W$ filters
during Cycle 6 (GO-6512; PI Hesser).  The observation log is described
in Table 1 of \citet[][hereafter D08]{do08a}.  We retrieved the AM 1 images from the STScI
archive and performed independent PSF fitting photometry.  The PSF was
modelled on each image using a number (between 40 and 80) of isolated
and well sampled stars, adopting the PSF/DAOPHOTII routine. A first
PSF fitting was then performed on every image using DAOPHOTII/ALLSTAR,
while ALLFRAME was used to obtain a fine measure of the star
magnitudes.  For each star all the magnitudes were normalized to a
reference frame and averaged together, and the photometric error was
derived as the r.m.s. of the repeated measurements.

For homogeneity with the S99 reduction procedure, we used the colour terms
in table 7 from \citet[][]{ho95} to convert the instrumental ($F555W,
F814W$) magnitudes to the standard Johnson V and Kron-Cousins I
(hereafter V and I, respectively).

%%%%%%%%%%%%%%%%%%%%%%%%%%%%%%%%%%%%%%% SECTION
%%%%%%%%%%%%%%%%%%%%%%%%%%%%%%%%%%%%%%% FIGURE
\begin{figure}
\centering
\includegraphics[scale=0.44]{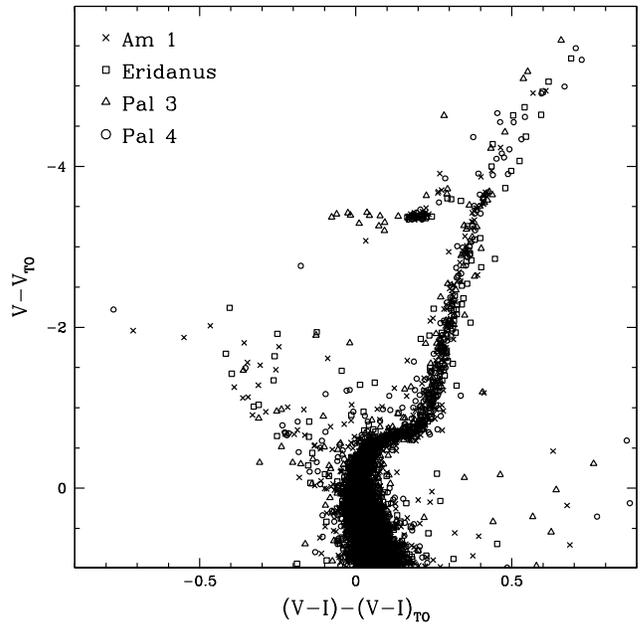}
\caption{The (V,V-I) CMDs of AM 1 (crosses), Eridanus (open squares),
  Pal 3 (open triangles), and Pal 4 (open circles) superimposed one to
  the other and aligned at the TO colour and magnitude, $(V-I)_{TO}$
  and $V_{TO}$, respectively.}
\label{fig_cmd_all}
\end{figure}
%%%%%%%%%%%%%%%%%%%%%%%%%%%%%%%%%%%%%%% FIGURE 

\subsection{Colour-magnitude diagrams and age}
\label{sec_cmd}
The CMDs obtained for AM 1, Eridanus, Pal 3 and Pal 4 are shown in
Figure \ref{fig_cmd} (see also figures 1--3 in S99).  All the detected
stars are shown as black dots, while the location of the mean ridge
line obtained with a second order polynomial interpolation with a 2.5
sigma clipping rejection criterion is displayed with grey points.  The high quality of
the WFPC2 images allows us to properly sample the clusters' stars from
the Tip of the Red Giant Branch (RGB) down to $\sim3$ mag below the
MS-TO.

The solid line in Figure \ref{fig_cmd} shows the location of the
best-fit theoretical isochrone from \citet[][]{do08b} over-plotted on
the CMDs of each cluster.  We used the distance modulus and reddening
from \citet[][2010 edition]{ha96}, while we adopted metallicity
[Fe/H]=--1.58 and --1.41 from \citet[][]{ko09} and \citet[][]{ko10} for
Pal 3 and Pal 4, respectively, [Fe/H]=--1.42 from S99 for Eridanus, and
[Fe/H]=--1.5 from D08 for AM 1.  With these parameters we obtain an
absolute age of 10.5 Gyr for Eridanus and Pal 4, and 11 Gyr for Pal 3
and AM 1, confirming that these clusters are
younger than the inner-halo GCs (S99 and D08).  

The location of the TO, defined as the hottest point along the MS in
the theoretical isochrone of each cluster, is also marked in Figure
\ref{fig_cmd}.  This was used to compute the shifts in magnitude and
colour that, after taking into account also the different distance
moduli and reddenings, are necessary to co-add the four CMDs.  The
result is impressive (see Figure \ref{fig_cmd_all}): the RGB
morphologies of the four clusters are in full agreement and consistent
with a small metallicity difference among their stellar populations.
Indeed, the combined CMD mimics a highly homogeneous
single-metallicity population.  These results are in agreement with the
study of S99 and D08, who find that these clusters are younger than M
3 by 1.5-2 Gyr, and they are homogeneous in terms of both age and
metallicity.
%% %%ha06%%%%%%%%%%%%%%%%%%%%%%%%%%%%%%%%%%%%% TABLE!!
 \begin{table}
 \caption{Center of gravity of the target clusters.}             % title of Table
 \label{tab_cent}      % is used to refer this table in the text
 \centering                          % used for centering table
 \begin{tabular}{l c c}        % centered columns (4 columns)
 \hline\hline                 % inserts double horizontal lines
 \noalign{\smallskip}
 Cluster & RA$_0$ & Dec$_0$  \\    % table heading 
 \noalign{\smallskip}
 \hline   
 \noalign{\smallskip}                     % inserts single horizontal line_
    AM 1  &    $03^{\rm h} 55^{\rm m} 02.5^{\rm s}$  &  $-49^\circ 36\arcmin 53.2\arcsec$     \\      
    Eridanus & $04^{\rm h} 24^{\rm m} 44.7^{\rm s}$   &  $-21^\circ 11\arcmin 13.9\arcsec$    \\
    Pal 3 &    $10^{\rm h} 05^{\rm m} 31.6^{\rm s}$  &  $00^\circ 04\arcmin 21.7\arcsec$        \\
    Pal 4 &    $11^{\rm h} 29^{\rm m} 16.5^{\rm s}$  &  $28^\circ 58\arcmin 22.4\arcsec$     \\
 \noalign{\smallskip}
 \hline                                   %inserts single line
 \end{tabular}
 \end{table}
%% %%%%%%%%%%%%%%%%%%%%%%%%%%%%%%%%%%%%%%% TABLE!!

\subsection{Astrometry and centre of gravity}
\label{sec_astro}

The coordinates of the stars identified in each cluster have been
converted from the ``local'' to the absolute astrometric system by
using a well tested procedure adopted by our group for more than 10
years \citep[see e.g.][and references therein]{la10}.  Since only a
few (if any) primary astrometric standards can be found in the central
region of GCs, we usually complement HST photometry with ground based
wide-field imaging.

B- and R-band images of AM 1 where taken in December 2000 with the Wide
Field Imager (WFI\footnote{The WFI is an imager composed of a mosaic
  of 8 CCD with a pixel scale of 0\farcs238/pix and a total FoV of
  $\sim34\arcmin\times33\arcmin$.}) at the MPE/ESO 2.2m telescope
(66.B-0454(A); PI: Gallager). Pal 3, Pal 4 and Eridanus were observed
in the $g$ and $r$ bands with MegaCam, the
$\sim1^{\circ}\times1^{\circ}$ FoV CCD camera at the
Canada-France-Hawaii Telescope (CFHT). We retrieved the bias-subtracted 
and flat-field corrected images from the CFHT Science Data
Archive (Observing run ID: 2010AC06 for Pal 3 and Pal 4; 2009BC02 for
Eridanus; PI Cote). We analysed the data following the procedure 
described in Section \ref{sec_data}. We analysed only the few chips (from 2 to 4,
depending on the location of the cluster centre in the FoV of each
data-set) which allow us to sample the entire cluster radial extent.
Notice that the tidal radius of the four clusters ranges between
$\sim2\arcmin$ and $\sim5\arcmin$ \citep[][]{ha96}. Unfortunately
the ground based observations reach only the TO level and were not
deep enough to properly sample the BSS region with an appropriate
signal-to-noise ratio. For this reason we used this data set only to
search for astrometric solutions and to supplement the HST data for the
construction of the cluster projected density profiles from resolved
star counts (see Sect. \ref{sec_dens}).
 
We used more than one hundred primary astrometric standard stars from
the GSC2.3 catalogue in the cluster vicinity to derive an astrometric
solution and obtain the absolute equatorial (RA and Dec)
positions of the stars
sampled in the ground based catalogues. Cross-correlations of the
catalogues and astrometric solutions were calculated with CataXcorr,
a code developed and maintained by Paolo Montegriffo at the INAF-Bologna
Astronomical Observatory \citep[see e.g.][]{la10,be11}.  Finally, the
positions of the stars in the WFPC2 data were transformed into the
same celestial coordinate system 
by adopting $\sim50$ stars in common
with the ground based catalogues.  The r.m.s. scatter of the final
solution was $\sim0.3\arcsec$ in both RA and Dec for the four
clusters.

The centre of gravity ($C_{grav}$) of the four clusters is estimated
as the barycenter of the resolved stars \citep[see e.g.][]{la10}. In
doing this an iterative procedure was adopted. The method estimates
the centre by averaging the RA and Dec positions of all the stars
contained within a circular area of a given radius and proceeds until 
convergence is reached.  We
performed this procedure using three different limiting radii and, at any
given radius, using only stars with magnitudes brighter than $V_{TO}$,
$V_{TO}+0.5$ mag and $V_{TO}+1$ mag. The final value of $C_{grav}$ is
calculated as the average of the nine measures (see Table
\ref{tab_cent}). The uncertainties both in RA and Dec are smaller than 1 arcsec for Eridanus and
AM 1, while they increase to $\sim 2\arcsec$ for Pal 3 and Pal 4 because of 
the extremely low stellar density even in the core region of these clusters.
Within the errors the new centres are in general good agreement with the ones listed by
~\citet[][]{ha96}.
%%%%%%%%%%%%%%%%%%%%%%%%%%%%%%%%%%%%%%% FIGURE
\begin{figure*}
\centering
\includegraphics[scale=0.44]{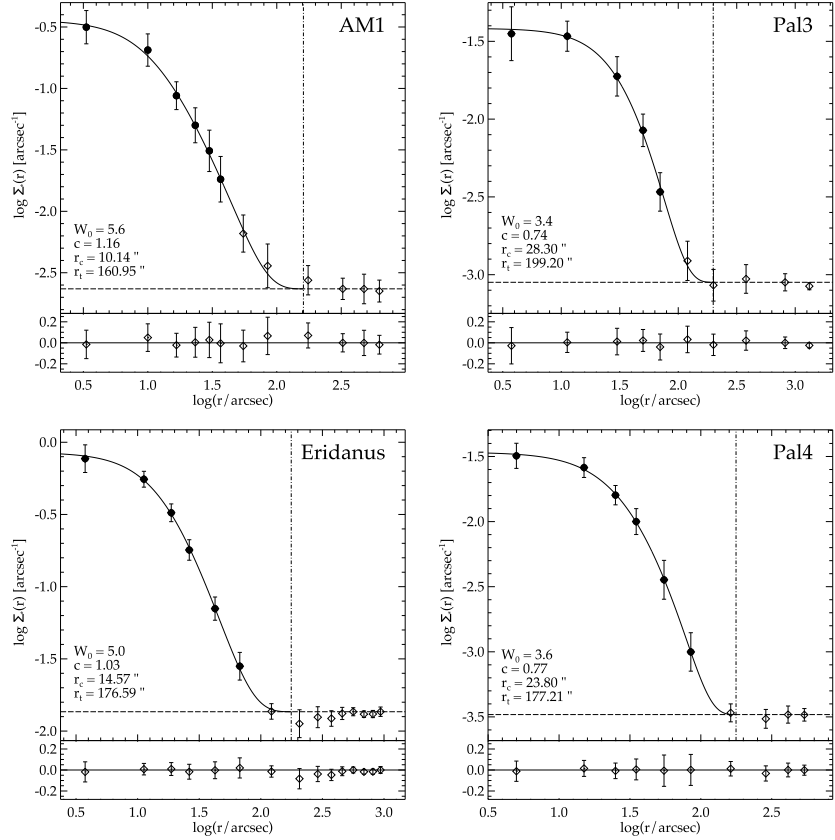}
\caption{Observed surface density profiles in units of number of stars
  per square arcsecond.  Star counts from the WFPC2 data are shown as
  solid circles, while those from ground based catalogues, normalised
  to the former, are shown as open diamonds.  The solid line
  represents the King model that best fits the observed density
  profile over the entire cluster extent.  Its characteristic
  parameters are labeled.  The dashed horizontal line shows the
  adopted stellar density background, while the dotted-dashed vertical
  line indicates the cluster tidal radius ($r_t$). The lower panel shows the residuals between 
  the observations and the fitted profile.}
\label{fig_prof}
\end{figure*}
%%%%%%%%%%%%%%%%%%%%%%%%%%%%%%%%%%%%%%% FIGURE

\section{Radial density profiles}
\label{sec_dens}
We used the photometric catalogues and the estimated values of
$C_{grav}$ to calculate the projected radial density profile from
resolved star counts for each cluster.  We divided the entire sample
into concentric annuli, each split into four subsectors
(quadrants).  The number of stars within each subsector was counted,
and the star density was obtained by dividing these values by the
corresponding subsector areas. The stellar density in each annulus was
then obtained as the average of the subsector densities.

The star counts in the clusters' central regions were performed using
the stars with a magnitude $V<(V_{TO}+1)$ from the WFPC2
high-resolution imaging catalogues, while the stars sampled down to
the TO were taken from the ground based data set for the external
regions.  Notice that the stars adopted to compute the density profile
cover a very limited range of mass in both the HST and the ground
based datasets, so that a single mass-model can be used to
  properly describe the observed profile (see next Section). The
final density profiles of the program clusters are shown in Figure
\ref{fig_prof}, where the star counts from ground based catalogues
(open diamonds) have been normalised to the WFPC2 ones (solid circles)
using at least two circular area in common between the two data sets.
These are the most accurate and extended radial density profiles ever
published for these GCs.
%############################################################# TABLE 
\begin{table*}
\tabletypesize{\scriptsize}
\begin{center}
\caption{Age and structural parameters of the four GCs. Ages were estimated
through the best-fit of theoretical isochrones. The uncertainties on the age are of the order of
0.2-0.3 Gyr.
  Cluster total
  masses have been computed from the observed total luminosities
  quoted by \citet{ha96} and the mass-to-light ratios estimated by
  \citet{ml05} (an average value of 1.9 has been adopted for the
  mass-to-light ratio of Eridanus, which is not included in this
  latter work).  The dimensionless potential, concentration, core and
  tidal radii and central mass density ($W_0$, $c$, $r_c$, $r_t$ and
  $\log(\rho_{0})$, respectively) are from the best-fit King
  models. The core relaxation time $\log(t_{rc})$ and half-mass
  relaxation time $\log(t_{rh})$ are computed following eq. (10) and
  (11) of \citet{dj93}, respectively.}
\label{tab_struc} % is used to refer this table in the text
\begin{tabular}{lcccccccccc}
\hline \hline
\noalign{\smallskip}
Cluster & Age & $\log(M_{cl})$ & $W_0$ & $c$ & $r_c$ & $r_t$ & $\log(\rho_{0})$ & $\log(t_{rc})$  & $\log(t_{rh})$ \\
        &[Gyr]&$[M_\odot]$ &       &     & [pc]  & [pc]  &$[M_\odot/\rm pc^3$]& [yr] & [yr] \\
\noalign{\smallskip}
\hline
\noalign{\smallskip}
AM1 &  11.0 & 4.10 & 5.60 & 1.16  &  5.99 & 95.12 & 0.48 & 9.02 & $9.55$\\
Eridanus & 10.5 & 4.26 & 5.00 & 1.03  &  6.35 & 77.01 & 0.06 & 9.32& $9.70$ \\
Pal3 & 11.0 & 4.48 & 3.40 & 0.74 & 12.74 & 89.72 & 0.04 & 9.95 & $9.95$\\
Pal4 & 10.5 & 4.62 & 3.60 & 0.77 & 12.59 & 93.77 & 0.18 & 9.96 & $10.01$\\
\noalign{\smallskip}
\hline
\end{tabular}
\end{center}
\end{table*}
%###############################################################################################

\subsection{King models and structural parameters}
\label{subsec_struc}
The derived density profiles allowed us to derive the clusters'
structural parameters through fitting isotropic King models \citep{ki66}.
The background density was estimated from the outer data
points.  The best fit has been determined by $\chi^2$ minimization.

As shown in Figure \ref{fig_prof}, the observed profiles are very well
reproduced by isotropic King models characterised by the quoted
parameters, with $W_0$, $c$, $r_c$ and $r_t$ indicating the central
dimensionless potential, the concentration, the core and tidal radii,
respectively \citep[see also Table \ref{tab_struc}; we adopted the
distances quoted by][]{ha96}.  These parameters have been used to
estimate the core relaxation time ($t_{rc}$) and the half-mass
relaxation time ($t_{rh}$) of the four clusters from equation (10) and
(11) of \citet{dj93}.  We assumed the average stellar mass $\langle
m_{\ast}\rangle = 0.42 M_{\odot}$~\citep[][]{so11} while the total
cluster masses $M_{cl}$ have been derived from the observed total
luminosity~\citep{ha96} and the mass-to-light ratios quoted by
\citet[][we adopted an average value of 1.9 for Eridanus, which is not
  included in that work]{ml05}.  The resulting values of $t_{\rm rc}$
and $t_{rh}$ are shorter than the age of the clusters
(Table \ref{tab_struc}). Hence, some degree of central mass segregation
is expected in the program clusters. This result is at odds with what
found in the two outer halo GCs previously investigated, NGC 2419 and
Pal 14, whose $t_{rh}$ ($\sim18$ and $\sim20$ Gyr, respectively) are
longer than the clusters' ages~\citep[$\sim12$ and $\sim 10.5$,
respectively; see][and D08]{da09}.
 %%%%%%%%%%%%%%%%%%%%%%%%%%%%%%%%%%%%%%% FIGURE
\begin{figure}
\centering
\includegraphics[scale=0.44]{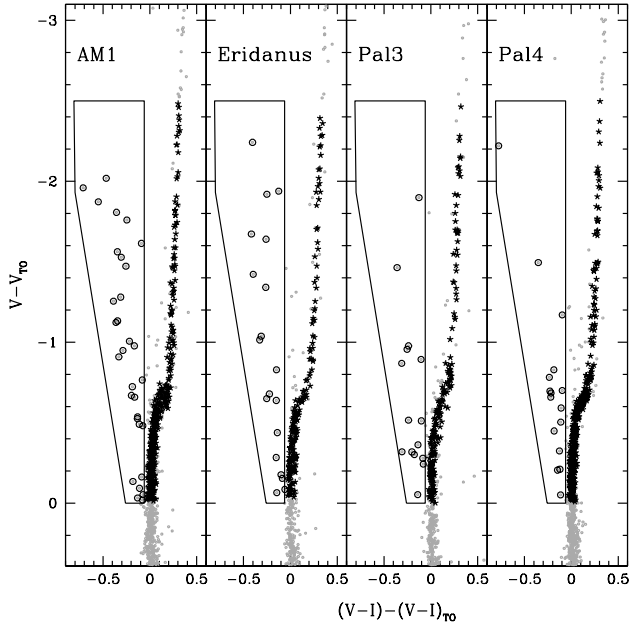}
\caption{Population selection in the four clusters. The bona-fide
  BSSs, located within the marked selection boxes, are shown as open
  circles. The reference population of RGB stars is marked with
  asterisks.}
\label{fig_sel}
\end{figure}
%%%%%%%%%%%%%%%%%%%%%%%%%%%%%%%%%%%%%%% FIGURE

\section{The BSS radial distribution}
\label{sec_bss}
As demonstrated by previous results \citep[see e.g.][and references
  therein]{fe04, car11, sa12,sal12}, BSSs in most Galactic GCs are
more centrally concentrated than normal stars. Since BSSs are more
massive than the average, and since the half-mass relaxation time in
those systems is much smaller than their age, this result is generally
ascribed to the effect of dynamical friction and mass segregation
\citep[e.g.][]{ma06}.

The clusters studied in this paper contain a significant number of
candidate BSSs \citep[see Figure \ref{fig_cmd_all}; see
also][hereafter S05]{san05}.  In order to investigate their radial
distribution, the BSS populations have been selected in a homogeneous
way in all clusters.  We assumed the TO as the reference point and
selected as bona-fide BSSs all the stars brighter than $V_{TO}$ and
$3\times\sigma_{VI}$ bluer than $(V-I)_{TO}$, where $V_{TO}$ and
$(V-I)_{TO}$ are, respectively, the magnitude and colour of the TO as
defined in Section \ref{sec_cmd}, while $\sigma_{VI}$ is the combined
photometric uncertainty in the colour. The adopted selection boxes and
the resulting bona-fide BSS samples are shown in Figure \ref{fig_sel}.

According to \citet{fe93}, in order to study the radial distribution
of BSSs, it is necessary to define a reference population which is
expected to follow the cluster light distribution.  Since the number
of horizontal branch stars in these GCs is very low, we decided to
adopt the RGB as the reference population. We selected as giants
all the stars in the range of magnitude $V_{TO}>V>V_{TO}-2.5$,
lying at a distance smaller than 3$\sigma$ from the mean ridge line
(see asterisks in Figure \ref{fig_sel}), where $\sigma$ is the uncertainty associated
with the mean ridge line (horizontal bars in Figure~\ref{fig_cmd}).  This choice allows us to obtain a populous
sample of reference stars in the same range of V magnitude, i.e., in
the same condition of completeness as for the BSSs.  The total number
of BSSs and RGB stars in each cluster is listed in Table
\ref{tab_count}.

%%%%%%%%%%%%%%%%%%%%%%%%%%%%%%%%%%%%%%% TABLE!!
\begin{table}
\caption{Number of BSSs and RGB stars selected in each cluster.}             % title of Table
\label{tab_count}      % is used to refer this table in the text
\centering                          % used for centering table
\begin{tabular}{l c c}        % centered columns (4 columns)
\hline\hline                 % inserts double horizontal lines
\noalign{\smallskip}
Cluster & N$_{\rm BSS}$ & N$_{\rm RGB}$  \\    % table heading 
\noalign{\smallskip}
\hline   
\noalign{\smallskip}                     % inserts single horizontal line_
   AM 1  &  31 & 352\\      
   Eridanus & 19 & 187   \\
   Pal 3 &  15 & 191    \\
   Pal 4 &  16 & 354   \\
\noalign{\smallskip}
\hline                                   %inserts single line
\end{tabular}
\end{table}
%%%%%%%%%%%%%%%%%%%%%%%%%%%%%%%%%%%%%%% TABLE!!

\subsection{Cumulative radial distribution and population ratios}
\label{sec_cumu}
\begin{figure}
\centering
\includegraphics[scale=0.44]{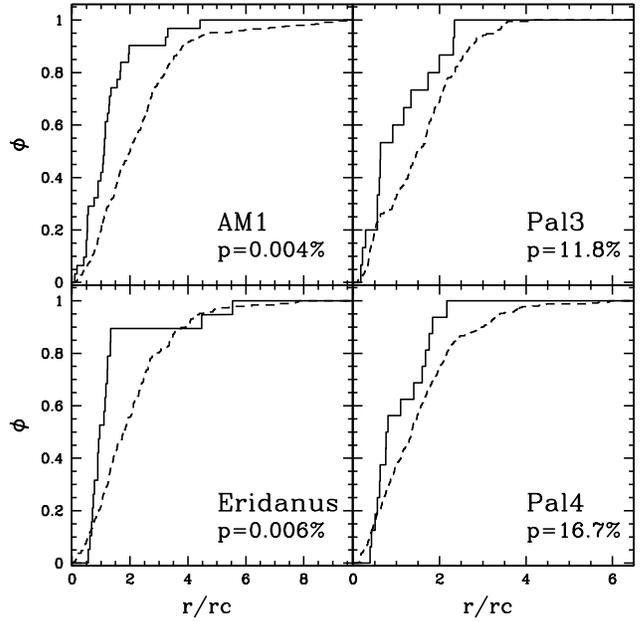}
\caption{Cumulative radial distributions of BBSs (solid lines) and RGB
  stars (dashed lines) as a function of their projected distance from
  the cluster centre normalised to the core radius ($r/r_c$), for each
  of the four clusters. The probability that the two populations are extracted
  from the same distribution is quoted in each panel.}
\label{fig_ks}
\end{figure}

In Figure \ref{fig_ks} we show the cumulative radial distribution of
BSSs (solid lines) and RGB stars (dashed lines) as a function of the
projected distance from the cluster centre ($r$) normalized to the
core radius $r_c$ estimated in Section \ref{subsec_struc}. A
  Kolmogorov-Smirnov test has been performed to check the probability
  that BSSs and RGB stars are extracted from the same parent
  distribution. Based on the probability values ($p$) obtained for
  each cluster and reported in Figure~\ref{fig_ks}, we conclude that
  BSS are more concentrated than the RGB in AM 1 and Eridanus, while
  the result is inconclusive for Pal 3 and Pal 4.

\begin{figure}
\centering
\includegraphics[scale=0.44]{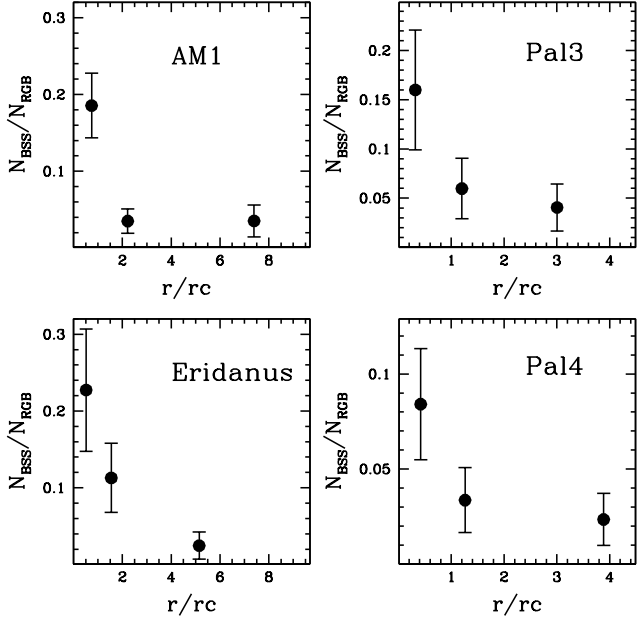}
\caption{Relative frequency of BSS-to-RGB stars, as a function of the
  projected distance from the centre normalized to the core radius
  ($r/r_c$) for each of the four clusters.}
\label{fig_rad}
\end{figure}

As second step to investigate the radial distribution of BSSs in the
four clusters, we calculated the ratio between the number of BSS
(N$_{\rm BSS}$) and that of RGB stars (N$_{\rm RGB}$), counted in
concentric annuli centred on the clusters' $C_{grav}$. The radial
distribution of N$_{\rm BSS}/{\rm N}_{\rm RGB}$ in the four clusters
is shown in Figure \ref{fig_rad}.  Consistently with what was found above,
BSSs appear to be systematically more centrally concentrated than RGB
stars. It is important to emphasise that, because of the insufficient
quality of the wide-field data, we are forced to limit our analysis to
the region sampled by the HST observations, which do not reach the
tidal radius of the program clusters. Thus, unfortunately we cannot
characterise the full shape of the BSS radial distribution, and we are
not able to conclude whether the distributions are bimodal or not.
However this analysis demonstrates that dynamical friction was already
effective in centrally segregating BSSs in the program clusters.

 %%%%%%%%%%%%%%%%%%%%%%%%%%%%%%%%%%%%%%% FIGURE
%%%%%%%%%%%%%%%%%%%%%%%%%%%%%%%%%%%%%%% FIGURE
%%%%%%%%%%%%%%%%%%%%%%%%%%%%%%%%%%%%%%% FIGURE
%%%%%%%%%%%%%%%%%%%%%%%%%%%%%%%%%%%%%%% FIGURE

\subsection{The Minimum Spanning Tree}
\label{sec_mst}
Given the relatively small number of BSSs, we applied the method of
the Minimum Spanning Tree \citep[MST; see][and references
  therein]{al09} as an alternative test to evaluate the degree of BSS
segregation.  The MST is the unique set of straight lines (``edges")
connecting a given sample of points (``vertices"; in this case the star
coordinates) without closed loops, such that the sum of the edge
lengths is the minimum possible.  Hence, the length of the MST is a
measure of the compactness of a given sample of vertices.
\citet{cw04} showed that the degree of mass segregation in a star
cluster can be measured by comparing the length of the MST of two
populations with different average masses \citep[see also][]{sce06}.
Notice that the MST method is independent of the catalogue astrometric
accuracy and of any assumption about the cluster center.  Moreover, it
gives quantitative measures of both the degree of mass segregation
and its associated significance.

Here we provide a detailed analysis of the segregation of BSSs with
respect to RGB stars in the four program clusters, based on a recent
modification of the MST method, where, instead of the direct sum of
$n$ edges of length $e_i$, their geometric mean is used \citep{ol11}:
\begin{equation}
\gamma_{\rm MST} = \left( \prod_{i=1}^n e_i \right)^{1/n} = \exp\left[
  \frac{1}{n} \sum_{i=1}^n \ln e_i \right].
\end{equation}
Following equation (2), we have computed $\gamma_{\rm MST}^{\rm mass}$
for the most massive population (i.e., the observed samples of $n+1$
BSSs in each cluster). Then, we randomly extracted $m$ sets of $n+1$
objects from the reference population (the combination of the BSS and
the RGB samples), and we computed their mean and standard deviation,
$\bar{\gamma}_{\rm MST}^{\rm ref}$ and $\Delta{\gamma_{\rm MST}^{\rm
    ref}}$ respectively. Finally, the level of BSS segregation with
respect to giant stars and its associated uncertainty have been
estimated as:
\begin{equation}
  %\begin{aligned}
    \Gamma_{\rm MST}         = \frac{\bar{\gamma}_{\rm MST}^{\rm ref}}{\gamma_{\rm MST}^{\rm mass}} \,,\\
    \Delta{\Gamma_{\rm MST}} = \frac{\Delta{\gamma_{\rm MST}^{\rm ref}}}{\gamma_{\rm MST}^{\rm mass}}.
  %\end{aligned}
\end{equation}
The geometrical mean has the very useful property of effectively
damping contributions from extreme edge lengths, so that $\Gamma_{\rm
  MST}$ of a compact configuration of even a few stars will not be
much affected by an ``outlier''. Note that $\gamma_{\rm MST}$ has the
dimension of a length, while $\Gamma_{\rm MST}$ is dimensionless.  A
value of $\Gamma_{\rm MST}=1$ is found if the two populations have the
same radial distribution, while $\Gamma_{\rm MST}>1$ indicates the
presence of mass segregation meaning a more concentrated distribution 
of massive stars compared to the reference population.  
In the latter case, the significance of mass segregation is provided by 
the value of $k$ for which $\Gamma_{\rm MST} - k\, \Delta{\Gamma_{\rm MST}} = 1$.

Table \ref{tab_mst} lists the values of $\Gamma_{\rm MST}$ and the
related $1\sigma$ uncertainties computed for the BSS populations in
the four target clusters,
and, for comparison, for NGC 2419 and Pal 14.  In agreement with the
results of \citet{da08} and \citet{bec11}, we find that BSS and RGB
stars share the same radial distribution in NGC 2419 and Pal 14
($\Gamma_{\rm MST}\simeq 1$).  Conversely, BSSs appear to be
significantly more segregated than giant stars in the other four
GCs. In particular, AM 1 and Eridanus show the highest level of mass
segregation (with $\Gamma_{\rm MST}\simeq 1.7$), with high statistical
significance ($2.5-3\sigma$). Pal 3 and Pal 4 show a slightly lower
degree of BSS segregation ($\Gamma_{\rm MST}\simeq 1.5$), at the $\sim
2\sigma$ level.

%%%%%%%%%%%%%%%%%%%%%%%%%%%%%%%%%%%%%%% TABLE!!
\begin{table}
\caption{Values of the MST tree ($\Gamma_{\rm MST}$) and its $1\sigma$
  uncertainty ($\Delta{\Gamma_{\rm MST}}$) for the BSS in the four target
  clusters and for NGC 2419 and Pal 14.}
  \centering
\begin{tabular}{l c c}        % centered columns (4 columns)
\hline\hline                 % inserts double horizontal lines
Cluster  & $\Gamma_{\rm MST}$& $\Delta{\Gamma_{\rm MST}}$ \\    % table heading 
\hline                        % inserts single horizontal line
AM 1     &             1.71 & 0.23 \\    
Eridanus &             1.69 & 0.27 \\    
Pal 3    &             1.56 & 0.27 \\    
Pal 4    &             1.53 & 0.29 \\    
NGC 2419 &             1.01 & 0.07 \\    
Pal 14   &             1.07 & 0.17 \\    
\noalign{\smallskip}
\hline                                   %inserts single line
\label{tab_mst}      % is used to refer this table in the text
\end{tabular}
\end{table}
%%%%%%%%%%%%%%%%%%%%%%%%%%%%%%%%%%%%%%% TABLE!!

\section{Summary}
\label{sec_summ}
Using HST-WFPC2 data we studied the BSS population in the central
regions of four Galactic halo GCs, namely AM 1, Eridanus, Pal 3 and
Pal 4.  These clusters, together with NGC 2419 and Pal 14, represent
the entire group of GCs at $R_{GC} > 50$ kpc.

A proper comparison of the derived CMDs with theoretical isochrones
confirms the younger ages (10.5-11 Gyr) of these systems with respect
to the inner-halo GCs (see Figure \ref{fig_cmd}). The impressive
similarity of their RGB morphology (Fig. \ref{fig_cmd_all}) also
indicates a small metallicity difference ($-1.41<$[Fe/H]$<-1.58$,
depending on the cluster), in agreement with previous findings by S99
and D08.

By complementing HST data with wide-field catalogues from ground based
imaging, we have derived the most extended radial density profiles from
resolved star counts ever published for these clusters
(Fig. \ref{fig_prof}). All profiles are well fit by isotropic King
models with the structural parameters listed in Table \ref{tab_struc}.

Unluckily, the insufficient quality of the wide field catalogues did
not allow us to study the BSS populations in the cluster outskirts and
look for possible signatures of an external rising branch, similar to
that found in most of the previously surveyed GCs \citep[see][ and
  references therein]{sa12}.  We therefore limited the analysis to the
area covered by the HST data.  BSSs have been selected in a homogenous
way in the four program clusters and their radial distribution has
been compared to that of RGB stars, taken as a reference.  We found
that BSSs are significantly more centrally concentrated than giants in
all four systems (see Figs.  \ref{fig_ks}, \ref{fig_rad} and Table
\ref{tab_mst}).  Since BSSs are assumed to be more massive than normal
cluster stars, their higher central concentration is interpreted in
terms of an effect of mass segregation. Indeed, the half-mass
relaxation time of the four program GCs is found to be smaller than
their age (see Table \ref{tab_struc}), thus indicating that dynamical
friction has been already effective in segregating the most massive
stars in the cores. By measuring the degree of mass segregation with
the $\Gamma_\mathrm{MST}$ test, we conclude that AM 1 and Eridanus are
the dynamically oldest systems in the group, while Pal 3 and Pal 4 are
slightly less dynamically evolved, and NGC 2419 and Pal 14 show no
signature of mass segregation yet \citep[see also][]{da08,bec11}.
The results shown in this paper once more indicate
that, indeed, BSSs represent the ideal population to investigate the
dynamical state of dense stellar systems.

\acknowledgments

The authors thank the anonymous referee for her/his useful
comments on the first version of the paper.
This research is part of the project COSMIC-LAB funded by the European
Research Council (under contract ERC-2010-AdG-267675).  The research
leading to these results has received funding from the European
Community's Seventh Framework Programme (/FP7/2007-2013/) under grant
agreement No 229517.  Based on observations made with the NASA/ESA
Hubble Space Telescope, obtained from the data archive at the Space
Telescope Institute. STScI is operated by the association of
Universities for Research in Astronomy, Inc. under the NASA contract
NAS 5-26555.  This research used the facilities of the Canadian
Astronomy Data Centre operated by the National Research Council of
Canada with the support of the Canadian Space Agency.
C.O. appreciates funding by the German Research Foundation (DFG), grant OL 350/1-1.
N.L thanks Holger Baumgardt for his great help to understand and apply the King models used in this paper.
%% To help institutions obtain information on the effectiveness of their
%% telescopes, the AAS Journals has created a group of keywords for telescope
%% facilities. A common set of keywords will make these types of searches
%% significantly easier and more accurate. In addition, they will also be
%% useful in linking papers together which utilize the same telescopes
%% within the framework of the National Virtual Observatory.
%% See the AASTeX Web site at http://www.journals.uchicago.edu/AAS/AASTeX
%% for information on obtaining the facility keywords.

%% After the acknowledgments section, use the following syntax and the
%% \facility{} macro to list the keywords of facilities used in the research
%% for the paper.  Each keyword will be checked against the master list during
%% copy editing.  Individual instruments or configurations can be provided 
%% in parentheses, after the keyword, but they will not be verified.

{\it Facilities:} \facility{ESO (WFI)}, \facility{HST (WFPC2)}.

%% Appendix material should be preceded with a single \appendix command.
%% There should be a \section command for each appendix. Mark appendix
%% subsections with the same markup you use in the main body of the paper.

%% Each Appendix (indicated with \section) will be lettered A, B, C, etc.
%% The equation counter will reset when it encounters the \appendix
%% command and will number appendix equations (A1), (A2), etc.

\end{document}